# From Silver Nanoparticles to Thin Films: Evolution of Microstructure and Electrical Conduction

*Haoyan Wei and Hergen Eilers*[*]

Applied Sciences Laboratory, Institute for Shock Physics, Washington State University, Spokane, WA 99210, USA

**ABSTRACT**

Silver nanoparticles embedded in a dielectric matrix are investigated for their potential as broadband-absorbing optical sensor materials. This contribution focuses on the electrical properties of silver nanoparticles at various morphological stages. The electrical current through thin films, consisting of silver nanoparticles, was characterized as a function of film thickness. Three distinct conductivity zones were observed. Two relatively flat zones ("dielectric" for very thin films and "metallic" for films thicker than 300 – 400 Å) are separated by a sharp transition zone where percolation dominates. The dielectric zone is characterized by isolated particle islands with the electrical conduction dominated by a thermally activated tunneling process. The transition zone is dominated by interconnected silver nanoclusters – a small increase of the film thickness results in a large increase of the electrical conductivity. The metallic conductivity zone dominates for thicknesses above 300 – 400 Å.

# 1. INTRODUCTION

Nanostructured silver particles exhibit unique optical characteristics. In contrast to their corresponding bulk counterparts, metallic nanoparticles can absorb electromagnetic radiation, resulting in surface plasmon polaritons at the metal-dielectric interface. The resonance wavelength of metallic nanoparticles is strongly dependent on the metal and the particle size and particle shape [1]. As initially isolated metallic nanoparticles form a fractal structure during the deposition process, the interconnection leads to nanoclusters with more irregular shapes and broader size distributions [2]. Such fractal structures can greatly extend the absorption from the visible wavelength band into the infrared (IR) wavelength region [3]. By adjusting the particle size and shape distribution, it is possible to tailor the optical properties of these materials to specific applications such as surface enhanced Raman scattering (SERS) [4, 5], color filters [6-8], and all optical switching [9]. Our interest in exploring the effects of microstructure and applied external electrical fields on the conduction of silver thin films near the percolation threshold is due to their potential use as multispectral optical sensors, that are sensitive to visible and infrared radiation [2].

We recently reported on the tailored plasmonic behavior of Ag/Teflon® AF nanocomposite materials. We were able to show a large broadband visible to infrared absorption spectrum [2] suitable for multispectral sensor applications. Also, we demonstrated a nanocomposite with an absorption spectrum that closely matched the solar radiation spectrum [10]. Westphalen et al. and Tian et al. showed that the excitation of surface plasmons in metal clusters can lead to the generation of photoelectrons [11, 12]. Pillai et al. showed that solar cells containing metallic nanoparticles can dramatically enhance the near infrared absorption due to the presence of surface plasmons [13].

The synthesis of such nanocomposites typically involves the deposition of metallic nanoparticles into a dielectric matrix. The electrical properties of such metal/dielectric nanocomposites are determined by the embedded metallic nanostructures [14]. Thus, the model to describe the electrical properties of discontinuous metal films on a dielectric substrate can be used to describe the conduction mechanism in metal/dielectric nanocomposites [14-16]. The metallic film structures can be described as a system of metallic nanoparticles embedded in a combined dielectric medium of glass substrate and vacuum spacing between particles.

Generally, the electrical conductivity of metallic films can be divided into three zones. Due to the isolated nature of discontinuous particles, extremely thin films (dielectric zone) show a very low conductivity. As the film thickness increases, the electrical conductivity rapidly increases as isolated particles start to coalesce (percolation zone). At film thicknesses close to the electron mean free path (EMFP), the film exhibits near metallic conductivity (metallic zone).

## 2. EXPERIMENTAL DETAILS

Silver was deposited onto glass and silicon substrates. Plain soda-lime glass microscope slides (75 x 25 x 1 mm$^3$) were purchased from Fisher Scientific and cut in half. Prime grade Si (100) wafers with a diameter of 2" were obtained from Silicon Quest International, cut into small squares about 8 x 8 mm$^2$, and used for scanning electron microscopy (SEM) imaging. Both types of substrates were sonicated in methanol and rinsed with copious amounts of deionized water before use. Silver wires of 99.999% purity (metal basis) were obtained from Alfa Aesar and used as is.

Silver films were deposited at room temperature by vapor-phase deposition in a high-vacuum chamber (base pressure of 10$^{-7}$ torr) equipped with a four-pocket electron-beam evaporator manufactured by Mantis Deposition Ltd. An Inficon quartz crystal microbalance (QCM) was used to gauge the deposition rate and film thickness. Different substrates have different silver condensation coefficients (ratio of the number of adsorbed metal atoms to the total number of metal atoms arriving at the substrate surface). Thus the QCM was calibrated with atomic force microscopy (AFM) measurements on a continuous silver film (thickness larger than 500 Å). Specimens for electrical measurements were deposited onto glass substrates because of their smooth surface and dielectric nature. Two co-planar metal wire electrodes were glued with water based conductive graphite adhesives, 10 mm apart, on the substrate surface (see the schematic representation in the inset in Fig. 4). In this two-probe configuration electrical measurements were performed at room temperature under vacuum condition using a Keithley 2400 source-meter unit (SMU) interfaced with a computer for automatic data collection. The SMU current and voltage ranges are 10 pA - 1.05 A and 1 µV - 210 V, respectively.

To characterize the morphology of thin silver films by SEM and to reduce charging during the microstructure imaging, it was necessary to deposit silver films onto silicon substrates. The surface of the silicon substrates is covered with a layer of amorphous native oxides (0.6-2.0 nm thick [17]) and is thus similar to the surface of the glass slides used for electrical tests. The silicon and glass substrates were placed side-by-side and coated in parallel.

Film thickness measurements were performed on a Molecular Imaging PicoPlus AFM in contact mode, using standard Si probes obtained from NanoAndMore USA. Film surface morphology was examined by an FEI Sirion 200 field emission SEM (FESEM) operating at 15 keV. ImageJ analysis software was used to determine the area filling fractions and particle sizes.

The film thicknesses were determined by QCM measurements and a constant calibration factor was applied to the measured data. This calibration factor was determined by comparing the QCM results of several samples with the corresponding AFM measurements. We would like to point out that this calibration factor is expected to depend on the film morphology and is thus not a constant.

## 3. RESULTS AND DISCUSSION

Fig. 1 shows a series of FESEM images of silver films at various thicknesses deposited at a constant rate of 0.067 Å/s. As the thickness increases, a transition from isolated island particles to interconnected clusters, and finally to a complete continuous film is observed. Typically, metals do not wet dielectric surfaces and their growth obeys the Volmer-Weber mechanism (island growth) [18]. At the initial deposition stage the metal atoms tend to coalesce into individual isolated islands to minimize the surface free energy. Fig. 1a shows individual isolated islands with a nominal film thickness of 17 Å. The particles are not uniform in size and their size distribution is shown in Fig. 2a. The particles have an average size of about 6-7 nm in diameter (assuming a round shape) and an area filling fraction $p$ of about 35%. As the deposition continues, the islands grow in all three dimensions. The growth in lateral direction leads to a decrease of the gap between islands. In Fig. 1b, the 50 Å thick film has a grain size of about 10-12 nm and a filling fraction of about 45%. For the film of 83 Å thickness in Fig. 1c, the grain size is about 17 nm and the filling fraction is about 62%. The corresponding distributions of particle sizes and interparticle spacings are plotted in Fig. 2 and indicate a log-normal distribution, in agreement with literature reports [19, 20]. As the films grow thicker, the size of the metal islands increases and the size distribution becomes broader.

The increasing size distribution indicates coalescence effects as neighboring particles get close to each other. The pairing and chaining of islands observed in Fig. 1b and Fig. 1c confirm this tendency. The coalescence leads to the formation of islands with irregular shapes resulting in fractal structure growth. Interconnected fractal clusters start to appear for the films shown in Fig. 1d and Fig. 1e (161 Å and 200 Å) in thickness respectively). Multiple, previously isolated islands form larger irregular shaped clusters, leading to a quasi-continuous film with filling fractions of about 73% and 82%, respectively. Fig. 1f shows a 259 Å thick film with further increased connectivity. Multiple, fully continuous electrical paths have been formed across the film. The substrate is almost completely covered with silver resulting in a high filling fraction of about 92%, which is already beyond the percolation threshold $p_c$. As the silver film continues to grow thicker, the morphology starts to resemble its bulk counterpart (Fig. 1g).

Although the QCM is calibrated with AFM measurements, it is more appropriate to consider the QCM readings as nominal, because the QCM converts the measured mass into a thickness measurement based upon the assumption of a uniform continuous configuration. The discontinuous nature of the film formed in the early deposition stage leads to a larger discrepancy in QCM readings due to the Volmer-Weber type growth. The relation between the filling fraction and the film thickness at a constant deposition rate of 0.067 Å/s was plotted in Fig. 3 where the data points exhibit an asymptotic trend.

Fig. 4 shows the measured normalized current in the silver film as a function of film thickness under various applied external voltages. As the applied voltage increases, the transition from the dielectric zone to the metallic zone is shifted to smaller film thicknesses. Because the contribution to the measured current from silver vapors present in the chamber could not be neglected for discontinuous films thinner than 100 Å, the current was measured after the residual metal vapors had disappeared and with the two existing shutters (blocking of e-gun and blocking of substrate) closed. In addition, the e-gun was turned off to avoid potential electromagnetic interference especially on low current measurement. Above 200 Å, when the film is percolated, the effect of the metal vapor on the current measurement is unperceivable and the current through the film was continuously monitored during the deposition.

The electrical conductivity of silver thin films can be divided into three zones. For very thin film (< 50-80 Å), it exhibits dielectric properties. Very low electrical currents are measured due to the relatively large separation of the discontinuous film nature (Fig. 1a, b and c). As the film thickness increases, a rapidly increasing current is observed in the transition zone where the dominant morphological features are interconnected islands resulting in a percolated network (Fig. 1d, e and f). At thicknesses larger than 300-400 Å, the resulting films exhibit metallic electrical properties.

For filling fractions $p < p_c$, the film is quasi-continuous, containing arrays of islands and clusters separated from each other. The separation distance typically ranges from 1 nm to 10 nm (Fig. 2b). The electrical conductivity of such quasi-continuous films is a complicated process which involves various conduction mechanisms including thermionic emission, thermally activated tunneling, metal conduction, and substrate conduction [21]. The dominating conduction mechanism is closely associated with the island film geometry and is determined by the film thickness. For films with islands separated by 1 - 10 nm, thermally activated tunneling is believed to be the dominating conduction mechanism [22-24]. For electrical conduction to occur, electrons have to be able to traverse from one island to the next across the island gap. This process requires the thermal activation energy to overcome the potential difference between the two islands [23].

The potential difference between the particles is the work that has to be done during charge transport from one particle to the other (see energy diagram in reference [18, 21]). Without an applied electric field, this Coulomb activation energy $U_a$ is related to the particle geometry by [23, 25]:

$$U_a = \frac{e^2}{4\pi\varepsilon_0\varepsilon_r}\left(\frac{1}{r} - \frac{1}{r+\Delta}\right)$$

where $\varepsilon_0$ is the permittivity of vacuum, $\varepsilon_r$ is the relative permittivity, $r$ is the island radius and $\Delta$ is the inter-island separation. For very thin films the separation is large and the particle size is small. Both, the potential barrier height and the activation energy are large. As the film thickness increases, the islands grow larger and their separation decreases (Fig. 2). This will not only lower the activation energy but also reduce the potential barrier width and height due to the proximity effect (overlap of image force potentials), leading to the increase of electron conductivity. In addition, as coalescence takes place, the total number of particles and gaps decrease and the conduction mechanism changes from mostly tunneling between islands to mostly metallic within the merged islands. The increase in electrical conductivity as the films grow is therefore attributed to the combined roles of (i) the eased condition for thermally activated tunneling (decreased activation energy, barrier height and width) and (ii) the gradually increased contribution from metallic conduction. When the islands are large enough and the gaps are small enough a substantial increase in current occurs, leading to the onset of the transition zone.

As the applied electrical field increases, the onset of the transition zone shifts both upward and leftward as shown in Fig. 4. The applied field affects the conduction by: (i) providing more electrons that can drift along the field direction; (ii) reducing the activation energy required to transport an electron by $\Delta e E_s$ ($E_s$ is the field between particles) [23], and (iii) lowering the parabolic potential barrier height (see the energy diagram in reference [18]), increasing the chance for electron tunneling to occur. Thus the effect of a decreased activation energy, and a decreased barrier height and width due to an applied electric field is equivalent to that of the film thickness buildup as discussed above. All these factors will contribute to the early onset of the transition zone under elevated external potentials.

Fig. 5 illustrates how well Ohm's law is obeyed in silver thin films with different thicknesses. Nonlinear behavior was observed in samples with film thicknesses below the percolation threshold. At present no definite conclusions have been drawn for the sub-linear deviation of extremely thin films less than 50 Å (Fig. 5a). It might originate from the charging effect of distantly separated islands which could act as capacitors.[26]

For samples with film thicknesses in the range of 50 Å to 170 Å (Fig. 5b), the conductance increases as the applied potential is increased. Similar super-linear effects were also observed in Ni, Pt and Au thin films [18, 23]. Several models have been proposed to explain the super-linear deviation from Ohm's law. One possibility is the variation of activation energy due to: (i) since the activation energy is closely related to the particle size, the log-normal size distribution of particles leads to a distribution of activation energies, which is broader for thicker films than for thinner films (Fig. 2); (ii) as the interparticle spacing becomes smaller, the interactions between charge carriers become stronger, which could impact both, the activation energy and the charge mobility [23], and (iii) as aforementioned, the application of an external electric field reduces the activation energy required for charge carrier creation in the field direction by $\Delta eE_s$. In addition, the applied electric field significantly reduces the height and width of the potential barrier at higher potential, increasing the probability for electron tunneling to occur.

Another model to explain the super-linear deviation from Ohm's law is the non-equilibrium heating of the electron gas [18]. Electron lattice scattering is negligible for nanoparticles as their size is smaller than the electron mean free path. Therefore the electron energy loss due to volume phonon generation is greatly reduced, which could favorably generate "hot" electrons with sufficient external energy input such as an applied potential. At elevated electron temperature, the maximum contribution to the tunneling is attributed to these hot electrons which are located at higher energy levels and closer to the barrier top. Thus they experience lower barrier heights and narrower widths (i.e. more transparency). Whatever the mechanism may be, the energy input from the external field is expected to enable more electrons to participate in the contribution of the induced current.

Fig. 4 also shows the non-ohmic electrical resistance for film thicknesses below 170 Å. Above 170 Å, the resistances measured for different voltages start to coincide, indicating the transition of electrical conduction from electron tunneling to metallic. The switch to metallic conduction indicates that a continuous electrical path has formed across the film as shown in Fig. 1e. The connectivity of the metallic particles continues to increase as the films continue to grow (Fig. 1f). At film thicknesses around 300-400 Å and thicker the observed electrical current and resistance becomes flat and the metallic zone starts. The conductivity is limited by the so called size effect [27], which is intrinsically linked to the electron mean free path (EMFP). The EMFP of silver is about 520 Å at room temperature [27]. At film thicknesses below the EMFP, the limiting factors for the electrical conductance are attributed to surface and grain boundary scattering [28].

Above 1000 Å (Fig. 1g), grain dimensions are large enough to exceed the EMFP and the size effect (surface and grain boundary scattering) is negligible. The electrical conductivity in this zone resembles the electrical conductivity of bulk silver with electron-phonon scattering being the dominant mechanism.

The midpoint in the transition region, which occurs at about 160-180 Å, is considered to be the percolation threshold with a corresponding metal filling fraction of about 70%-80% It appears that the percolation threshold occurs at the film thickness where the resistance deduced under different voltages starts to coincide with each other. This occurrence is expected since the percolation threshold is defined as the point where the first electrical pathway across the film is formed and the change from activated tunneling to metallic conduction as the dominant conduction mechanism occurs.

Fig. 6 shows the effect of the deposition rate on the electrical current through the film in the transition zone. At same film thicknesses, lower conductances were observed for films synthesized at low deposition rates than those synthesized at high deposition rates. This difference stems from the different microstructures formed during deposition as indicated in Fig. 1e and Fig. 7. More continuous paths (superimposed white lines) without encountering gaps and narrow necks (black boundary lines between clusters) between particles are found for films synthesized at higher rates, indicating an increased connectivity between nanoclusters. As aforementioned, the non-wetting nature of metal atoms on dielectric surfaces leads to their Volmer-Weber growth which is a diffusion-limited nucleation and growth process. Thus, lower deposition rates allow metal atoms more time to migrate and coalesce into more isolated particles. Higher deposition rates lead to a steeper slope in the transition zone, thus reaching the metallic zone at smaller film thicknesses.

The film growth can also be affected by other deposition parameters such as substrate temperature [29] and the choice of deposition method (sputtering [30], energetic condensation [31] and laser ablation [16]). Changes in the silver mobility can lead to distinct electrical and optical properties. Such effects may benefit the fine tuning of silver nanoparticles for different applications. At present, our interest focuses on potential application of silver nanoparticles for multispectral optical sensors and photovoltaic cells based on surface plasmon excitation.

## 4. CONCLUSION

The electrical conductance of silver films was investigated as a function of film thickness and under various applied external electrical fields. Three conductivity zones with distinct microstructures were observed: dielectric (film consists of isolated particle islands), transition (film consists of percolated metallic network), and metallic (film consists of a metallic continuum). The electron transport in the dielectric zone is governed by an activated tunneling process, while the electron transport in the metallic zone can be described by Ohm's law. The transition zone between the dielectric and the metallic zones is characterized by the appearance of the first conductive pathway across the film. An external electrical field applied across the film can shift the onset of this transition zone to lower film thicknesses. This shift is due to the change of activation energy. The slope in the transition zone can be tuned by varying the deposition rate. Such metal/dielectric systems have potential applications in surface plasmon mediated multispectral sensors, photovoltaic modules, and optical/electrical switches.

## ACKNOWLEDGMENTS

This work was supported by ARO grant W911NF-06-1-0295 and by ONR Grant N00014-03-1-0247.

**REFERENCES**

[1]     Y.N. Xia, N.J. Halas, Shape-controlled synthesis and surface plasmonic properties of metallic nanostructures, MRS Bulletin 30 (2005), 338-344.

[2]     A. Biswas, H. Eilers, F. Hidden, O.C. Aktas, C.V.S. Kiran, Large broadband visible to infrared plasmonic absorption from Ag nanoparticles with a fractal structure embedded in a Teflon AF (R) matrix, Appl. Phys. Lett. 88 (2006), 013103.

[3]     V.M. Shalaev, Optical Properties of Nanostructured Random Media, Springer, Berlin, 2002.

[4]     D.M. Kuncicky, B.G. Prevo, O.D. Velev, Controlled assembly of SERS substrates templated by colloidal crystal films, J. Mater. Chem. 16 (2006), 1207-1211.

[5]     M. Moskovits, Surface-enhanced Raman spectroscopy: a brief retrospective, J. Raman Spectrosc. 36 (2005), 485-496.

[6]     A. Biswas, O.C. Aktas, U. Schurmann, U. Saeed, V. Zaporojtchenko, F. Faupel, T. Strunskus, Tunable multiple plasmon resonance wavelengths response from multicomponent polymer-metal nanocomposite systems, Appl. Phys. Lett. 84 (2004), 2655-2657.

[7]     Y. Dirix, C. Bastiaansen, W. Caseri, P. Smith, Oriented pearl-necklace arrays of metallic nanoparticles in polymers: A new route toward polarization-dependent color filters, Adv. Mater. 11 (1999), 223-227.

[8]     M. Quinten, The color of finely dispersed nanoparticles, Appl. Phys. B: Lasers Opt. 73 (2001), 317-326.

[9]     G.I. Stegeman, E.M. Wright, All-Optical Wave-Guide Switching, Opt. Quantum Electron. 22 (1990), 95-122.

[10]    H. Eilers, A. Biswas, T.D. Pounds, M.G. Norton, M. Elbahri, Teflon AF/Ag nanocomposites with tailored optical properties, J. Mater. Res. 21 (2006), 2168-2171.

[11]    M. Westphalen, U. Kreibig, J. Rostalski, H. Luth, D. Meissner, Metal cluster enhanced organic solar cells, Sol. Energy Mater. Sol. Cells 61 (2000), 97-105.

[12]    Y. Tian, T. Tatsuma, Mechanisms and applications of plasmon-induced charge separation at TiO2 films loaded with gold nanoparticles, J. Am. Chem. Soc. 127 (2005), 7632-7637.

[13]    S. Pillai, K.R. Catchpole, T. Trupke, M.A. Green, Surface plasmon enhanced silicon solar cells, J. Appl. Phys. 101 (2007), 093105.

[14]    A. Kiesow, J.E. Morris, C. Radehaus, A. Heilmann, Switching behavior of plasma polymer films containing silver nanoparticles, J. Appl. Phys. 94 (2003), 6988-6990.

[15]    K. Seal, M.A. Nelson, Z.C. Ying, D.A. Genov, A.K. Sarychev, V.M. Shalaev, Metal coverage dependence of local optical properties of semicontinuous metallic films, J. Mod. Opt. 49 (2002), 2423-2435.

[16]    K. Seal, M.A. Nelson, Z.C. Ying, D.A. Genov, A.K. Sarychev, V.M. Shalaev, Growth, morphology, and optical and electrical properties of semicontinuous metallic films, Phys. Rev. B 67 (2003), 035318.

[17]    H.F. Okorn-Schmidt, Characterization of silicon surface preparation processes for advanced gate dielectrics, IBM J. Res. Dev. 43 (1999), 351-365.

[18]     R.D. Fedorovich, A.G. Naumovets, P.M. Tomchuk, Electron and light emission from island metal films and generation of hot electrons in nanoparticles, Phys. Rep. 328 (2000), 73-179.

[19]     J.A. Blackman, B.L. Evans, A.I. Maaroof, Analysis of Island-Size Distributions in Ultrathin Metallic-Films, Phys. Rev. B 49 (1994), 13863-13873.

[20]     P.G. Borziak, Y.A. Kulyupin, S.A. Nepijko, V.G. Shamonya, Electrical conductivity and electron emission from discontinuous metal films of homogeneous structure, Thin Solid Films 76 (1981), 359-378.

[21]     R.M. Hill, Electrical Conduction in Ultra Thin Metal Films. I. Theoretical, Proc. R. Soc. London, A 309 (1969), 377-395.

[22]     J.E. Morris, Recent developments in discontinuous metal thin film devices, Vacuum 50 (1998), 107-113.

[23]     C.A. Neugebauer, M.B. Webb, Electrical conduction mechanism in ultrathin, evaporated metal films, J. Appl. Phys. 33 (1962), 74-82.

[24]     E. Dobierzewska-Mozrzymas, E. Pieciul, P. Bieganski, G. Szymczak, Conduction mechanisms in discontinuous Pt films, Cryst. Res. Technol. 36 (2001), 1137-1144.

[25]     P. Bieganski, E. Dobierzewska-Mozrzymas, E. Pieciul, G. Szymczak, Influence of microstructure on the conduction mechanisms in discontinuous metal films on dielectric substrates, Vacuum 74 (2004), 211-216.

[26]     T. Ohgi, H.Y. Sheng, Z.C. Dong, H. Nejoh, D. Fujita, Charging effects in gold nanoclusters grown on octanedithiol layers, Appl. Phys. Lett. 79 (2001), 2453-2455.

[27]    W. Zhang, S.H. Brongersma, O. Richard, B. Brijs, R. Palmans, L. Froyen, K. Maex, Influence of the electron mean free path on the resistivity of thin metal films, Microelectron. Eng. 76 (2004), 146-152.

[28]    W. Wu, S.H. Brongersma, M. Van Hove, K. Maex, Influence of surface and grain-boundary scattering on the resistivity of copper in reduced dimensions, Appl. Phys. Lett. 84 (2004), 2838-2840.

[29]    B. Gergen, H. Nienhaus, W.H. Weinberg, E.M. McFarland, Morphological investigation of ultrathin Ag and Ti films grown on hydrogen terminated Si(111), J. Vac. Sci. Technol., B 18 (2000), 2401-2405.

[30]    M. Arbab, The base layer effect on the d.c. conductivity and structure of direct current magnetron sputtered thin films of silver, Thin Solid Films 381 (2001), 15-21.

[31]    E. Byon, T.W.H. Oates, A. Anders, Coalescence of nanometer silver islands on oxides grown by filtered cathodic arc deposition, Appl. Phys. Lett. 82 (2003), 1634-1636.

**List of Figure Captions**

Fig. 1. SEM micrographs show the transition of silver from isolated individual islands to interconnected networks as the film thickness increases. The thickness of the films is 17 Å (a), 50 Å (b), 83 Å (c), 161 Å (d), 200 Å (e), 259 Å (f), and >5000 Å (g), respectively. The films were deposited at a rate of 0.067 Å/s except (g), which was deposited at a rate of 1.5 Å/s.

Fig. 2. The distribution of particle size (a) and interparticle spacing (b) of silver islands as illustrated in Fig. 1a, b and c. The log-normal curves are superimposed.

Fig. 3. The change of silver filling fraction with film thickness and deposition time at a constant deposition rate of 0.067 Å/s.

Fig. 4. The normalized electrical current and resistance of silver films on a glass substrate as a function of film thickness under various applied voltages. The film deposition was kept at a constant rate of 0.067 Å/s.

Fig. 5. Log of conductance vs. the square root of applied external field at different film thickness. Deviations from Ohm's law are observed for discontinuous films (Figs. 5a and 5b).

Fig. 6. The dependence of the normalized electrical current on the deposition rate. A steeper slope was observed at higher deposition rates.

Fig. 7. Morphology of 200 Å thick silver films at different deposition rates. (a) 0.67 Å/s and (b) 2.0 Å/s. Increased connectivity is observed for films with higher deposition rates. Only one continuous path in (a) while three in (b).

**FIGURES**

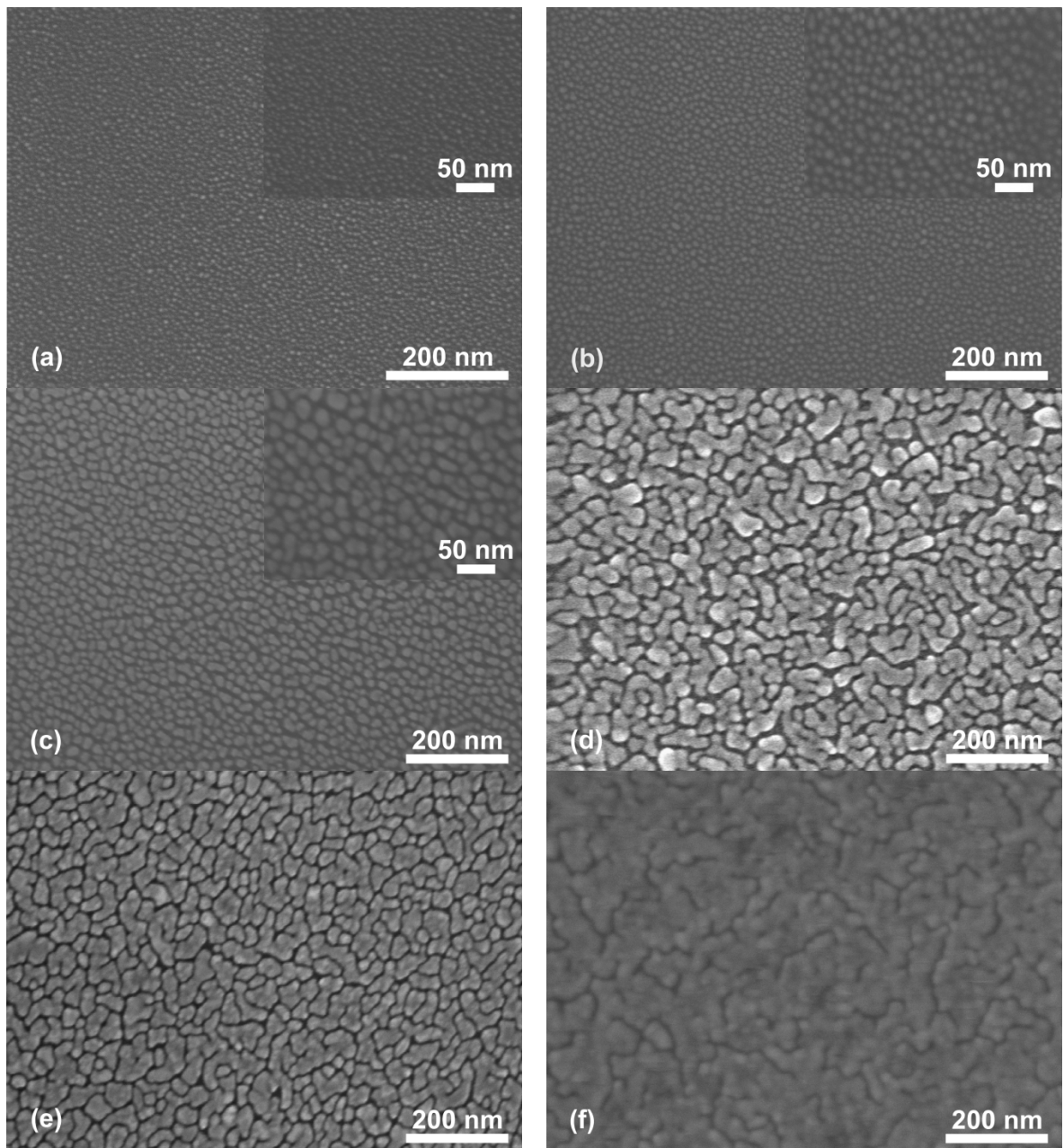

Fig 1

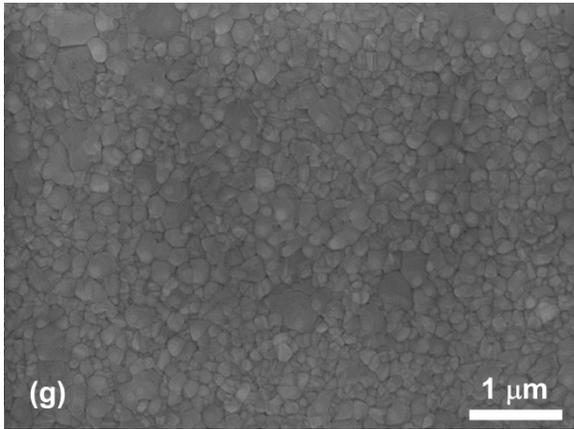

Fig. 1 continuation

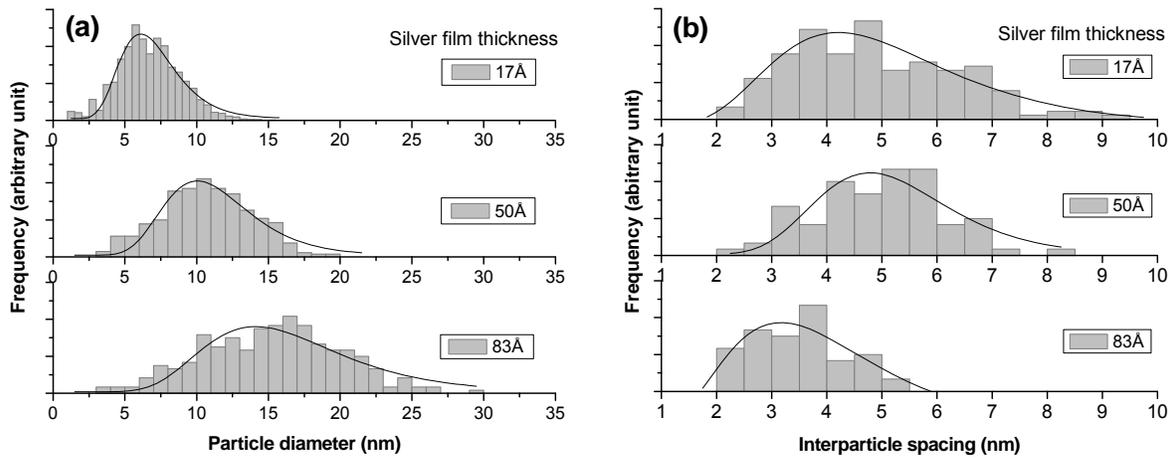

Fig. 2.

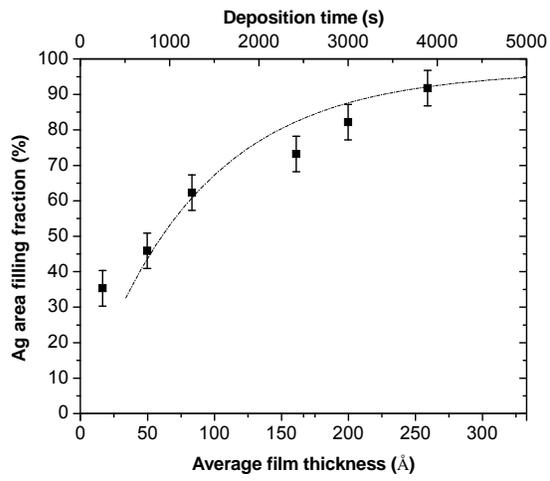

Fig. 3.

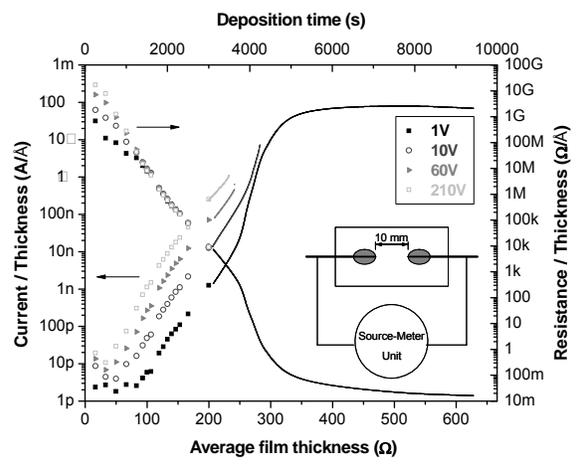

Fig. 4.

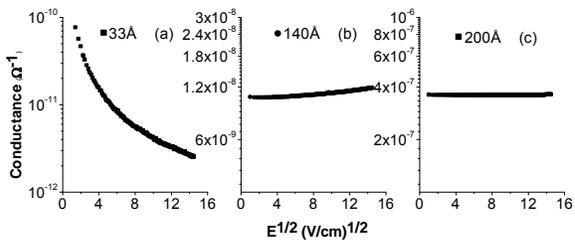

Fig. 5.

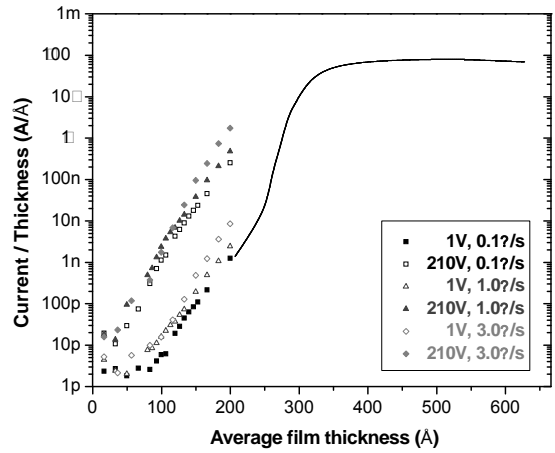

Fig. 6.

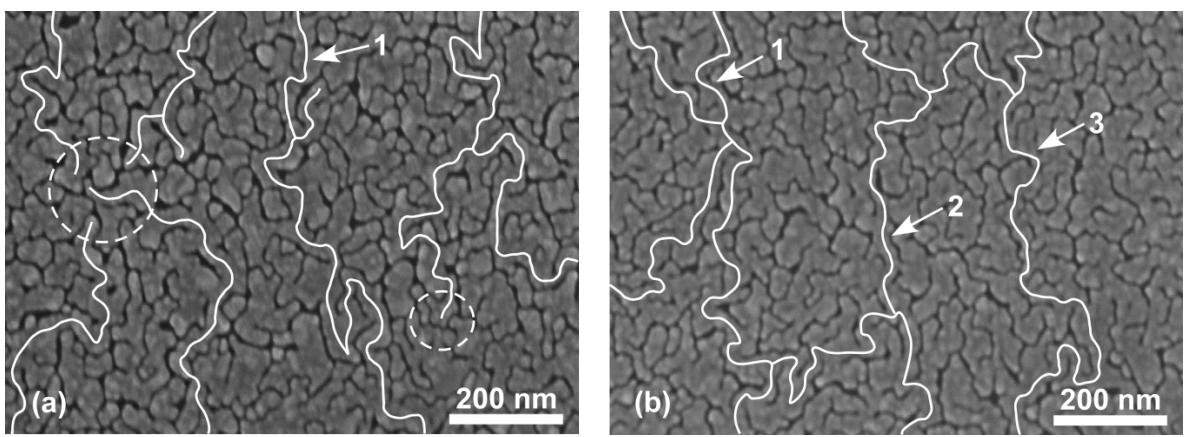

Fig. 7.